\documentstyle[12pt,epsfig]{article}
\begin{document}
 \newcommand{\beq}{\begin{equation}}\newcommand{\eeq}{\end{equation}}
 \newcommand{\barr}{\begin{eqnarray}}\newcommand{\earr}{\end{eqnarray}}
\newcommand{\andy}[1]{ }

 \def\tltl{\widetilde}

 \def\bmn{\mbox{\boldmath $n$}} \def\bmA{\mbox{\boldmath $A$}}
 \def\bmB{\mbox{\boldmath $B$}} \def\bmb{\mbox{\boldmath $b$}}
 \def\bmsigma{\mbox{\boldmath $\sigma$}}
 \def\bmsigman{\mbox{\boldmath $\sigma$}\cdot\mbox{\boldmath $n$}}
 \def\bmsigmab{\mbox{\boldmath $\sigma$}\cdot\mbox{\boldmath $b$}}
 \def\bmsigmaA{\mbox{\boldmath $\sigma$}\cdot\mbox{\boldmath $A$}}
 \def\bmsigmaB{\mbox{\boldmath $\sigma$}\cdot\mbox{\boldmath $B$}}

 \def\ch{\mbox{ch}}
 \def\sh{\mbox{sh}}
 \newcommand{\ket}[1]{| #1 \rangle}
 \newcommand{\bra}[1]{\langle #1 |}

 \def\NI{\noindent} \def\ellip{$\ldots$}
 \def\undertext#1{$\underline{\hbox{#1}}$}
 \def\lsnote#1{\def\dash{\hbox{\rm---}}{\bf~[[}~{\tt #1~$\ldots\,$LS}{\bf]]~}}
 \def\asteriskbreak{\vbox{\medskip\hrule\vskip -11pt
    $$************************************************$$\vskip-3pt
    \hrule\medskip}}
 \def\beginbignote{\begingroup\baselineskip 13pt\tt\def\dash{\hbox{\rm---}}
     \bigskip\hrule\bigskip}  \def\endbignote{\bigskip\hrule\bigskip\endgroup}
 \def\nl{\hfil\break}    \def\eqn#1{Eq.\ (\ref{eq:#1})}
 \def\pp{{\vec p}} \def\AA{{\vec A}} \font\romsix=cmr6 scaled\magstep0
 \def\coltwovector#1#2{\left({#1\atop#2}\right)}
 \def\up{\coltwovector10}    \def\down{\coltwovector01}
 \def\header#1{{
 \removelastskip\vskip 20pt plus 40pt \penalty-200 \vskip 0pt plus -32pt
 \NI\bf #1}\nobreak\medskip\nobreak}
 \font\romeight=cmr8 scaled\magstep0  \font\boldeight=cmbx8 scaled\magstep0
 \font\italeight=cmti8 scaled\magstep0
 \def\ask{\marginpar{?? ask:  \hfill}}
 \def\fin{\marginpar{fill in ... \hfill}}
 \def\lyrics#1{{\bf~[[LYRICS:}~{\bf #1}~{\bf]]}}
 \def\note{\marginpar{$\bigg|$ note \hfill}}
 \def\check{\marginpar{check \hfill}}
 \def\discuss{\marginpar{discuss \hfill}}

 \begin{titlepage}
 \begin{flushright}
 \today \\
BA-TH/99-334\\
 \end{flushright}
 \vspace{.5cm}
 \begin{center}
{\LARGE Berry phase from a quantum Zeno effect } 
\\ \quad

{\large P. FACCHI,$^{(1)}$ A.G. KLEIN,$^{(2)}$ \\
S. PASCAZIO$^{(1)}$ and L. S. SCHULMAN$^{(3)}$\\
           \quad    \\

        $^{(1)}$Dipartimento di Fisica, Universit\`a di Bari \\
and Istituto Nazionale di Fisica Nucleare, Sezione di Bari \\
 I-70126  Bari, Italy

        $^{(2)}$School of Physics, The University of Melbourne \\
Parkville, Victoria, Australia 3052

        $^{(3)}$Physics Department, Clarkson University \\
 Potsdam, NY 13699-5820, USA \\and\\Physics Department, Technion, Haifa, Israel

}

 \vspace*{.5cm}
 \end{center}
 \NI PACS: 03.65.Bz; 03.75.Be; 03.75.Dg


 \vspace*{.5cm}


 \begin{center}{\small\bf Abstract}\\ \end{center}

{\small We exhibit a specific implementation of the creation of
geometrical phase through the state-space evolution generated by the
dynamic quantum Zeno effect.  That is, a system is guided through a
closed loop in Hilbert space by means a sequence of closely spaced
projections leading to a phase difference with respect to the original
state. Our goal is the proposal of a specific experimental setup in which
this phase could be created and observed. To this end we study the case
of neutron spin, examine the practical aspects of realizing the
``projections," and estimate the difference between the idealized
projections and the experimental implementation.} \bigskip

 \end{titlepage}

 \newpage

 \setcounter{equation}{0}
 \section{Introduction }
 \label{sec-introd}
 \andy{intro}

The effect of the observer in quantum mechanics is perhaps nowhere more
dramatic than in the collection of phenomena loosely (and casually) known
as the ``quantum Zeno effect." This was first formulated by von Neumann
\cite{von,Beskow}, and is deeply rooted in fundamental features of the
temporal behavior of quantum systems \cite{strev}. During the last decade
there has been much interest in this issue, mainly because of an idea due
to Cook \cite{Cook}, who proposed using two-level systems to check this
effect, and the subsequent experiment performed by Itano {\em et al.}
\cite{Itano1}. New experiments were proposed, based on the physics of the
simplest of two-level systems: Neutron spin and photon polarization
\cite{qze1,inn}. 

Most of the referenced papers deal with what might be called the 
``static" version of the quantum Zeno effect. However, the most
striking action of the observer is not only to stop time 
evolution (e.g., by repeatedly checking if a system has decayed), 
but to {\it guide} it. In this article we will be concerned with a 
``dynamical" version of the phenomenon: we will show how guiding a 
system through a closed loop in its state space (projective 
Hilbert space) leads to a geometrical phase 
\cite{Panchar,BerryQuantal,BerryClassical,Shapere,Wagh}. This was 
predicted on general grounds \cite{AA87}, but here we use a 
specific implementation on a spin system \cite{continuous} and 
propose a particular experimental context in which to see this 
effect. It is remarkable that the Berry phase that is discussed 
below is due to measurements only: no Hamiltonian is needed. 

\setcounter{equation}{0}
\section{Forcing the pot to boil }
\label{sec-potboil}
\andy{potboil}

 We summarize the main features of the quantum Zeno effect (QZE). Prepare a
quantum system in some initial state $\psi(0)$. In time $dt$, by the
Schr\"odinger equation, its phase changes by $\hbox{O}(dt)$ while the
absolute value of its scalar product with the initial state changes by
$\hbox{O}(dt^2)$.

 The {\it dynamical\/} quantum Zeno effect exploits the above features and
forces the evolution in an arbitrary direction by a series of repeated
measurements: Let $\psi$ evolve with the Hamiltonian $H$, so that in the
absence of observations its evolution would be $\psi(T)=\exp(-iHT)
 \psi(0)$ (we take $\hbar=1$ throughout). Let there be a family of states
$\phi_k$, $k=0,1,\ldots, N$, such that $\phi_0=\psi(0)$, and such that
successive states differ little from one another (i.e.,
$|\langle\phi_{k+1} | \phi_k \rangle|$ is nearly 1). Now let $\delta T =
T/N$ and at $T_k=k\delta T$ project the evolving wave function on
$\phi_k$. Then for sufficiently large $N$, $\psi(T) \approx \phi_{_N}$.
[The usual QZE is the special case $\phi_k=\phi_0 (=\psi(0)) \ \forall \
k$.]

 In the following we consider an experiment involving a neutron spin. It
should be clear, however, that our proposal is valid for any system with
the same two-level structure.

\subsection{Evolution with no Hamiltonian}
\label{sec-noH}
\andy{noH}

 Assume first that there is {\em no} Hamiltonian acting on the system: one
can think, for instance, of a neutron crossing a region where no magnetic
field is present. The time-evolution is due to measurement only.

 The system starts with spin up along the $z$-axis and is projected on the
family of states
 \andy{projfamily}
 \beq
 \phi_k \equiv \exp(-i\theta_k\bmsigman)\up  \qquad
   \hbox{with~} \theta_k \equiv \frac{ak}N  \;,
  \qquad k=0,\ldots,N \ ,
 \label{eq:projfamily}
 \eeq
 where $\bmsigma$ is the vector of the Pauli matrices and $\bmn =
(n_x,n_y,n_z)$ a unit vector (independent of $k$).

 We assume that the system evolves for a time $T$ with projections at
times $T_k = k\delta T$ ($k=1,\dots,N$ and $\delta T=T/N$). The final
state is $\left[\phi_0 = \up\right]$
 \andy{finstate}
 \barr
 \ket{\psi(T)}
  &=& |\phi_N\rangle \langle \phi_N|
       \phi_{N-1}\rangle \cdots \langle \phi_2|
       \phi_1\rangle \langle \phi_1| \phi_0\rangle \nonumber \\
  &=& |\phi_N\rangle
       \left(\cos \frac{a}{N} + i n_z \sin \frac {a}{N} \right)^N \nonumber \\
  &=& \cos^N \left(\frac{a}{N} \right)
       \left(1 + i n_z \tan \frac {a}{N} \right)^N
       |\phi_N\rangle  \nonumber \\
  &\stackrel{N\rightarrow \infty}{\longrightarrow} & \exp (ia n_z)
     |\phi_N\rangle  \nonumber \\
         & = & \exp (ia n_z) \exp (-ia\bmsigman) | \phi_0\rangle .
 \label{eq:finstate}
 \earr
 Therefore, as $N\to\infty$, $\psi(T)$ is an eigenfunction of the final
projection operator $P_N$, with unit norm. If $\cos\Theta \equiv n_z$ and
$a=\pi$, \andy{finstatepi}
 \beq
 \psi(T) = \exp (i \pi \cos \Theta) (-1) \phi_0
        = \exp [-i \pi (1-\cos \Theta)]  \phi_0
        = \exp (-i \Omega/2 )  \phi_0 ,
 \label{eq:finstatepi}
 \eeq
 where $\Omega$ is the solid angle subtended by the curve traced by the
spin during its evolution. The factor $ \exp (-i\Omega/2)$ is a Berry
phase and it is due only to measurements (the Hamiltonian is zero).
Notice that no Berry phase appears in the usual quantum Zeno context,
namely when $\phi_k \propto \phi_0 \ \forall \ k$, because in that case
$a=0$ in (\ref{eq:finstate}).

 To provide experimental implementation of the mathematical process just
described, one could (in principle) let a neutron spin evolve in a
field-free region of space. With no further tinkering, the spin 
state would not change. However, suppose we place spin filters 
sequentially projecting the neutron spin onto the states of 
\eqn{projfamily}, for $k=0,\ldots,N$. Thus the neutron spin is 
forced to follow another trajectory in spin space. The essence of 
the mathematical demonstration just provided is that while $N$ 
measurements are performed, the norm of wave function that is 
absorbed by the filters is $N\cdot$O$(1/N^2)=$O$(1/N)$. For 
$N\to\infty$, this loss is negligible. Meanwhile, as a result of 
these projections, the trajectory of the spin (in its space) is a 
cone whose symmetry axis is $\bmn$. By suitably matching the 
parameters, the spin state can be forced back to its initial state 
after time $T$ \cite{continuous}. 

It is interesting to look at the process (\ref{eq:finstate}) for $N$
finite. The spin goes back to its initial state after describing a
regular polygon on the Poincar\'e sphere, as in Figure 1a.
 \begin{figure}
 \begin{center}
\epsfig{file=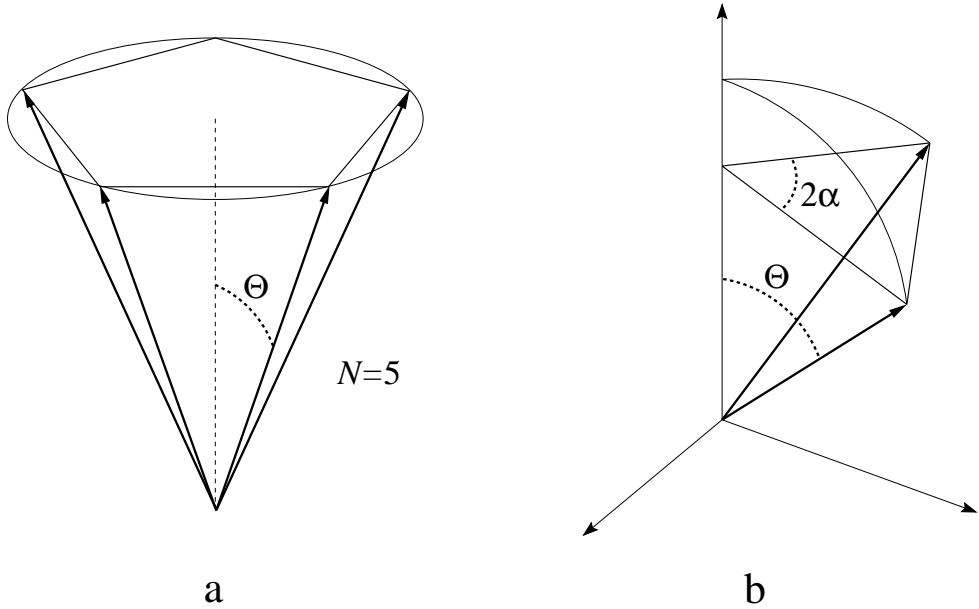, height=8cm}
 \caption{a) Spin evolution due to $N=5$ measurements.
b) Solid angles.}
 \end{center}
 \end{figure}
 After $N (<\infty)$ projections the final state is
 \beq
 \ket{\psi(T)}=\left(\cos\frac{a}{N}+in_z\sin\frac{a}{N}\right)^N
 \exp(-ia\bmsigman)\ket{\phi_0}.
 \eeq
 For $a=\pi$ the spin describes a closed path and
 \andy{clpath}
 \barr
 \ket{\psi(T)}&=&\left(\cos\frac{\pi}{N}+in_z\sin\frac{\pi}{N}\right)^N
 \exp(-i\pi)\ket{\phi_0}\nonumber\\
 &=&\left(\cos^2\frac{\pi}{N}+n^2_z\sin^2\frac{\pi}{N}\right)^{\frac{N}{2}}
 \exp\left(iN\arctan\left(n_z\tan\frac{\pi}{N}\right)\right)\exp(-i\pi)
 \ket{\phi_0}.\nonumber\\
 \label{eq:clpath}
 \earr
The first factor in the far r.h.s.\
accounts for the probability loss ($N$ is finite and there is
no QZE). We can rewrite (\ref{eq:clpath}) in the following form
 \beq
 \ket{\psi(T)}=\rho_N \exp(-i\beta_N)\ket{\phi_0},
 \eeq
 where
 \andy{rhoN, betaN}
 \barr
 \rho_N &=& \left(\cos^2\frac{\pi}{N}+n^2_z
 \sin^2\frac{\pi}{N}\right)^{\frac{N}{2}},
 \label{eq:rhoN}\\
 \beta_N &=& \pi-N\arctan\left(\cos\Theta\tan\frac{\pi}{N}\right).
 \label{eq:betaN}
 \earr
 In the ``continuous measurement" limit (QZE), we have
 \barr
 \rho &=& \lim_{N\to\infty}\rho_N=1,\nonumber\\
 \beta &=& \lim_{N\to\infty}\beta_N=\pi(1-\cos\Theta)=\frac{\Omega}{2},
 \earr
 where $\Omega$ is the solid angle subtended by the circular path, viewed
at an angle $\Theta$ (see Figure 1a). We recover therefore the result
(\ref{eq:finstatepi}).

 The relation between the solid angle and the geometrical phase is valid
also with a finite number of polarizers $N$. Indeed, it is
straightforward to show that the solid angle subtended by an 
isosceles triangle with vertex angle equal to $2\alpha$ (Figure 1b) 
has the value 
 \beq
 \Omega_{2\alpha}=2\alpha-2\arctan(\cos\Theta \tan\alpha).
 \eeq
 Hence if the polarizers are equally rotated of an angle $2\pi/N$, the
spin describes a regular $N$-sided polygon, whose solid angle is
 \beq
 \Omega_{(N)}=N\Omega_{2\pi/N}=2\pi-
2N\arctan\left(\cos\Theta\tan\frac{\pi}{N}\right)=2\beta_N,
 \eeq
 where we used the definition (\ref{eq:betaN}).
This result is of course in agreement with other analyses \cite{SM} 
based on the Pancharatnam connection \cite{Panchar}.

The above conclusion can be further generalized to the general case of an
arbitrary (not necessarily regular) polygon. Indeed, if the polarizers
are rotated at (relative) angles $\alpha_n$ with $n=0,\dots,N$, so that
 \beq
 \sum_{n=1}^N 2\alpha_n=2\pi,
 \eeq
 the solid angle is
 \beq
 \Omega'_{(N)}=\sum_{n=1}^N \Omega_{2\alpha_n}=
2\pi-2\sum_{n=1}^N\arctan(\cos\Theta \tan\alpha_n).
 \eeq 
 This is also twice the Berry phase. Notice that if all $\alpha_n\to0$
as $N\to\infty$ one again obtains the limit (\ref{eq:finstatepi}):
 \beq
 \Omega'=\lim_{N\to\infty}\Omega'_N
=2\pi-2\lim_{N\to\infty}\sum_{n=1}^N  \alpha_n \cos\Theta
=\Omega.
 \eeq

We emphasize that these predictions for the $N<\infty$ case are not 
trivial from the physical point of view. The above phases are 
computed by assuming that, during a ``projection" {\em \`a la} von 
Neumann, the spin follows a geodesics on the Poincar\'e sphere. The 
mathematics of the projection has no such assumptions. The 
``postulate's" only job is to relate all this projection formalism 
to measurements. 

 \subsection{Evolution with a non-zero Hamiltonian}
 \label{sec-yesH}
 \andy{yesH}

 Let us now consider the effect of a non-zero Hamiltonian
 \andy{Hamadd}
 \beq
H=\mu \bmsigmab ,
 \label{eq:Hamadd}
 \eeq
 where $\bmb = (b_x,b_y,b_z)$ is a unit vector, in general different from
$\bmn$. One can think of a neutron spin in a magnetic field. See Figure
2.
 \begin{figure}
 \begin{center}
\epsfig{file=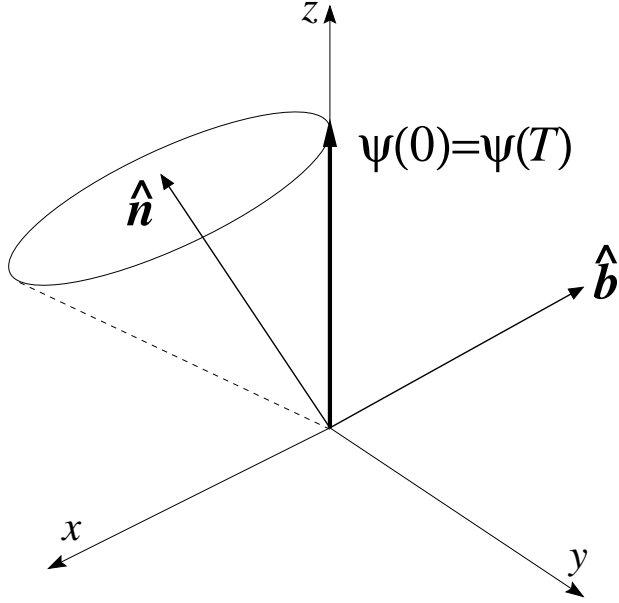, height=8cm}
 \caption{Spin evolution with measurements and non-zero Hamiltonian.}
\end{center}
 \end{figure}

If the system starts with spin up it would have the
following---undisturbed---evolution:
 \andy{undisturb}
 \beq
 \psi(t) = \exp(-i\mu t\bmsigmab)\up .
 \label{eq:undisturb}
 \eeq
  Now let the system evolve for a time $T$ with projections at times
$T_k=k\delta T$ ($k=1,\dots,N$ and $\delta T=T/N$) and Hamiltonian
evolution in between. Defining $P_0 \equiv |\phi_0\rangle \langle \phi_0|
= \pmatrix{1&0\cr0&0\cr}$, the $2\times2$ projection operator at
stage-$k$ is
  \andy{projk}
 \beq
P_k =|\phi_k\rangle \langle \phi_k|
 =\exp(-i\theta_k\bmsigman) P_0 \exp(i\theta_k\bmsigman)
 \label{eq:projk}
 \eeq
 and the state evolves to
 \andy{projev}
 \beq
 \psi(T)= \left[
 \prod_{k=1}^N \left[P_k \exp(-i\mu \delta T \bmsigmab)\right]\right]\up,
 \label{eq:projev}
 \eeq
where here and in subsequent expressions a time-ordered  product is
understood [with earlier times (lower $k$) to the right]. Using
$P_0^2=P_0$, Eq.\ (\ref{eq:projev}) can be rewritten
 \andy{projev2}
 \beq
 \psi(T)=  \exp(-ia\bmsigman) \left[ \prod_{k=1}^N  B_k \right] \up ,
 \label{eq:projev2}
 \eeq
with
 \andy{bk}
 \beq
 B_k \equiv
     P_0\exp(i\theta_k\bmsigman)
        \exp(-i\mu \delta T \bmsigmab)
            \exp(-i\theta_{k-1}\bmsigman)  P_0
 \label{eq:bk}
 \eeq ($\theta_0\equiv 0$).  The computation of $B_k$ requires a bit of
SU(2) manipulation. By using
 \andy{su21,2}
 \barr
 \left[ \bmsigmaA,\bmsigmaB \right] &=& 2i \bmsigma \cdot \bmA \times \bmB
 \label{eq:su21}   \\
(\bmsigmaA) (\bmsigmaB) (\bmsigmaA) &=& 2 (\bmA \cdot \bmB)\bmsigmaA
                                       - (\bmA\cdot\bmA)\bmsigmaB ,
 \label{eq:su22}
 \earr
valid for $c$-number $\bmA$ and $\bmB$, one gets
 \andy{sbA}
 \barr
 \exp(i\theta\bmsigman) \bmsigmab \exp(-i\theta\bmsigman)
             =\bmsigma\cdot \tltl{\bmb},
 \label{eq:sbA}
 \earr
 with
 \andy{sbB}
 \beq
 \tltl{\bmb}(\theta) \equiv
      \bmb \cos 2\theta + \bmn  (\bmb \cdot \bmn)(1-\cos2\theta)
       + \bmb \times \bmn \sin 2\theta ,
 \label{eq:sbB}
 \eeq
 which is the vector $\bmb$ rotated by $2\theta$ about the $\bmn$-axis.
The calculation of $B_k$ is now straightforward:
 \andy{sbk}
 \barr
 B_k &
   = & P_0\exp(i\delta \theta \bmsigman)
     \exp(-i\mu \delta T \bmsigma\cdot \tltl{\bmb}(\theta_{k-1})) P_0
                                                               \nonumber \\
    & = & P_0 \left(1+i\delta \theta \bmsigman
          -i\mu \delta T \bmsigma\cdot \tltl{\bmb}(\theta_k) \right) P_0
               + O(1/N^2),
 \label{eq:sbk}
 \earr
where $\delta \theta = \theta_{k+1} - \theta_k$ is $k$-independent.
Second order terms in $1/N$ drop out when the product (\ref{eq:projev2})
is computed for $N\to\infty$, so that
 \andy{prodB}
 \barr
 \prod_{k=1}^N  B_k
  &=& \prod_{k=1}^N P_0 (1+i\delta \theta \bmsigman -i \mu \delta T
       \bmsigma\cdot \tltl{\bmb}(\theta_k)) P_0                    \nonumber \\
  &=& \prod_{k=1}^N
     \left\{ P_0+i P_0 (\delta \theta \bmsigman - \mu \delta T
                \bmsigma\cdot \tltl{\bmb}(\theta_k)) P_0   \right\} \nonumber\\
  &=& \prod_{k=1}^N
      P_0\left\{1+i [\delta \theta n_z - \mu \delta T
                 \tltl{b}_z(\theta_k)] \right\}     \nonumber\\
 & = & P_0\exp\left\{  i \sum_{k=1}^N \left(
 \delta \theta n_z - \mu \delta T
                 \tltl{b}_z(\theta_k) \right)\right\}
  \label{eq:prodB}
 \earr
 where we have used $P_0 \sigma_x P_0= P_0 \sigma_y P_0=0$ and $P_0
\sigma_z P_0=P_0$. The continuum limit can be computed by letting the
summations in (\ref{eq:prodB}) become integrals in $dT$ and $d\theta$.
Moreover, $\frac{dT}{d\theta}=\frac{T}{a}$, which enables one to change
integration variable and get for the ``(1,1)" component of $\prod_{k=1}^N
B_k$ (all other components being zero)
 \andy{prodB11}
 \barr
 \exp \left( i n_z \int_0^a d \theta
    -i \mu \frac Ta
      \int_0^a \left[b_z \cos 2\theta
        + (\bmb \cdot \bmn) n_z (1- \cos 2 \theta)
          + (\bmb \times \bmn)_z \sin 2\theta \right] d\theta \right)
              \nonumber \\
 = \exp \left(i n_z a
    -i \mu \frac{T}{a} \left[ b_z \frac{\sin 2a}{2}
        + (\bmb \cdot \bmn) n_z \left(a- \frac{\sin 2a}{2} \right)
          + (\bmb \times \bmn)_z \frac{1-\cos 2a}{2} \right]\right),
             \nonumber \\
  \label{eq:prodB11}
 \earr
 The final state is an eigenstate of $P_N$ with unit norm, {\em
independent\/} of the Hamiltonian $H$:
 \andy{finpsi}
 \barr
 \psi(T) &=& \exp \left( - i\mu \frac{T}{a} \left[
b_z \frac{\sin 2a}{2} + (\bmb \cdot \bmn) n_z \left(a-\frac{\sin 2a}{2}
 \right) + (\bmb \times \bmn)_z \frac{1-\cos 2a}{2}
 \right]\right)
      \nonumber \\
 & & \times \exp\left(i a n_z  - ia\bmsigman \right) \up.
 \label{eq:finpsi}
 \earr
 The first factor in (\ref{eq:finpsi}) is obviously the ``dynamical
phase." Note that up to a phase, $\psi(t)$ is just $\phi_k$, with
$k=tN/T$. Therefore
 \andy{dynam}
 \barr
 \int_0^T \langle \psi(t) | H | \psi(t) \rangle dt
   &=& \frac{T}{a} \int_0^a \langle \phi_0|
       \exp(i\theta\bmsigman) \mu \bmsigmab
       \exp(-i\theta\bmsigman) |\phi_0\rangle d\theta
   \nonumber \\
   &=& \mu T
       \left[ b_z \frac{\sin 2a}{2a}
           + (\bmb \cdot \bmn) n_z \left(1 - \frac{\sin 2a}{2a} \right)
           + (\bmb \times \bmn)_z \frac{1-\cos2a}{2a} \right] , \nonumber\\
 \label{eq:dynam}
 \earr
 because the phases drop out in the above sandwich. It follows that the
remaining phase in (\ref{eq:finpsi}), when the spin goes back to its
initial state, is the geometrical phase.  When $a=\pi$
 \andy{fun}
 \beq
 \psi(T) = \exp \left( - i\Omega/2 \right)
 \exp\left(- i\mu T (\bmb \cdot \bmn) n_z \right) \up ,
 \label{eq:fun}
 \eeq
 where $\Omega$ is the solid angle subtended by the curve traced out by
the spin, as in (\ref{eq:finstatepi}), and $\mu T (\bmb \cdot \bmn) n_z$
yields the dynamical phase, as can also be seen by direct computation of
(\ref{eq:dynam}).
We remark that if time ordered products are looked upon as path integrals
\cite{PInt},
then our above demonstration is effectively a path integral derivation
of the geometrical phase.

 A practical implementation of the process just described would involve
an experimental setup similar to the one described after
\eqn{finstatepi}, but with a magnetic field whose action on the spin is
described by the Hamiltonian (\ref{eq:Hamadd}). If the neutron were to
evolve {\em only} under the action of the Hamiltonian, its spin would
precess around the magnetic field. However, the sequence of spin filters,
which project the neutron spin onto the states (\ref{eq:projfamily}),
compel the spin to follow the same trajectory as in the previous case
[Eq.\ (\ref{eq:finstate})], i.e.\ a cone whose symmetry axis is $\bmn$.
As above, the spin acquires a geometrical phase, but now there is a
dynamical phase as well.

 \subsection{A particular case}
 \label{sec-partc}
 \andy{partc}

It is instructive to look at a particular case of
(\ref{eq:finpsi})-(\ref{eq:fun}). We first note that if $\mu=0$ in
(\ref{eq:finpsi}) we recover (\ref{eq:finstate}). Now let $\bmb = \bmn$.
In this situation the projectors and the Hamiltonian yield the same
trajectory in spin space (although, as will be seen, at different rates).
If $\mu=0$ (so that $H=0$), the spin evolution is only due to the
projectors and the final result was computed in 
(\ref{eq:finstatepi}) 
 \andy{finbis}
 \beq
 \psi(T) = \exp (-i \Omega/2 )  \phi_0 .
 \label{eq:finbis}
 \eeq
 If, on the other hand, there is a nonvanishing Hamiltonian
(\ref{eq:Hamadd}), but {\em no} projectors are present, a cyclic
evolution of the spin is obtained for $\mu T=\pi$. The calculation is
elementary and yields
 \andy{fintris}
 \beq
 \psi(T) = \exp (-i \pi )  \phi_0 .
 \label{eq:fintris}
 \eeq
 Observe that the dynamical phase in this case is [$\mu T=\pi, \bmb =
\bmn$ and $a=\pi$ in Eq.\ (\ref{eq:dynam})]
 \andy{dynamss}
 \beq
 \int_0^T \langle \psi(t) | H | \psi(t) \rangle dt
 = \pi n_z = \pi[1- (1-n_z)] = \pi -\Omega/2 .
 \label{eq:dynamss}
 \eeq
 Therefore, the ``$\pi$" phase in (\ref{eq:fintris}) can be viewed, {\em
\`a la} Aharonov and Anandan \cite{AA87}, as the sum of a geometrical
($\Omega/2$) and a dynamical ($\pi - \Omega/2$) contribution.

 Now let both the Hamiltonian and the projectors be present. From Eq.\
(\ref{eq:fun}), one gets
 \andy{funny}
 \beq
 \psi(T) = \exp \left( - i\Omega/2 \right)
 \exp\left(- i\mu T n_z \right) \up ,
 \label{eq:funny}
 \eeq
 Notice that the value of $\mu$ is now arbitrary, so that $\mu T$ is not
necessarily equal to $\pi$ (the cyclic evolution of the spin is due 
to the projectors, not to the Hamiltonian). When $\mu T < \pi$, the 
projections are too ``fast" and do not yield (\ref{eq:fintris}). On 
the other hand, when $\mu T > \pi$, the projections are too slow 
and supply less phase, in comparison with Eq.\  (\ref{eq:fintris}). 
Only in the case  $\mu T = \pi$ do the projections yield the right 
phase in (\ref{eq:fintris}). Their presence is superfluous in this 
case: one would obtain exactly the same vector and the same phase 
without them. Our conclusions are summarized in Table~1. In 
some sense, one may say that the Hamiltonian dynamics provides a 
``natural clock" for the phase of the wave function. 
\begin{center}
{\small {\bf Table 1}: Phases for cyclic spin evolutions}  \\  \quad \\
\begin{tabular}{|c|c|c|c|}
  \hline\hline
      & $H=0$          & $H=\mu \bmsigmab$  & $H=\mu \bmsigmab$  \\
      & and projections & no projections      & and projections \\
  \hline
$\phi_{\rm geom}$ & $\Omega/2$ & $\Omega/2$ & $\Omega/2$ \\
\hline
$\phi_{\rm dyn}$  &  0 & $\pi - \Omega/2$ & $\mu T n_z$ \\
 \hline
$\phi_{\rm tot}=\phi_{\rm geom}+\phi_{\rm dyn}$ & $\Omega/2$ & $\pi 
(=\mu T)$ & $\Omega/2 + \mu T n_z$ \\ 
 \hline
& cyclic evolution &   cyclic evolution & cyclic evolution \\ & due 
to projections & due to field & due to projections \\ 
  \hline\hline
\end{tabular}
\end{center}

 \setcounter{equation}{0}
 \section{A Gedanken Experiment}
 \label{sec-expimpl}
 \andy{expimpl}

An  experimental implementation with neutrons would be difficult because
it would involve putting a QZE set-up inside an interferometer in order
to measure phase.  We therefore restrict ourselves to a gedanken
experiment based on the use of  $^3$He as a neutron polarization filter
\cite{Heil}.  It is well known \cite{Passel} that Helium 3 is ``black" to
neutrons but {\em polarized\/} $^3$He only absorbs one spin state of a
neutron beam---hence acts as a 50$\%$ absorber of a beam; the rest of it
emerges fully polarized.  In practice an external magnetic field is used
to maintain the polarization axis of the $^3$He.  If this external bias
field were to be given a slow twist along a longitudinal axis, the state
of polarization of the $^3$He should follow the direction of the twist.
A neutron beam propagating through a cell of high-pressure polarized
$^3$He along an axis aligned with the direction of twist will become
fully polarized and should develop a Berry phase according to the
argument of the previous section. 

From an experimental perspective a significant problem is that we 
so far lack a notion of slowness (as when we speak of ``slow twist" 
of the $B$ field). In the previous calculation, it is implicitly 
assumed that $\theta$ changes more slowly than $t$ (time): in other 
words, the relaxation processes in the $^3$He are given enough time 
(are fast enough) to function as a polarizer. A full treatment of 
this problem should therefore describe the physics of the 
projection process. We now tackle this issue and see that the 
notion of slowness can be given quantitative meaning in terms of a 
condition for adiabaticity. 

In practice, the absorption of the non-selected spin state occurs over a
finite distance, of the order of one or two centimeters. This situation
can be modeled via the following family of effective (nonhermitian)
Hamiltonians:
 \andy{effham}
 \beq
H_k = -i V |\phi_k^\perp \rangle \langle \phi_k^\perp |,
 \label{eq:effham}
 \eeq
where $V$ is a real constant and
 \andy{projfamily2}
 \beq
 \phi_k^\perp \equiv \exp(-i\theta_k\bmsigman)\down  \qquad
   \hbox{with~~} \theta_k \equiv \frac{ak}N  \;,
  \qquad k=0,\ldots,N \ .
 \label{eq:projfamily2}
 \eeq Note that $\langle \phi_k | \phi^\perp_k \rangle = 0$ [see Eq.\
(\ref{eq:projfamily})]. We first assume, for simplicity, that no external
($^3$He aligning) magnetic field is present. We define
 \andy{projk2}
 \beq
P_k^\perp \equiv |\phi_k^\perp \rangle \langle \phi_k^\perp|
 =\exp(-i\theta_k\bmsigman) P_0^\perp \exp(i\theta_k\bmsigman) \qquad
(P_0^\perp = |\phi_0^\perp \rangle \langle \phi_0^\perp|) \ .
 \label{eq:projk2}
 \eeq Obviously $P_k^\perp = 1-P_k$, where $P_k$ was defined in
(\ref{eq:projk}). The evolution engendered by the above Hamiltonian reads
 \andy{effhamev}
 \beq
e^{-iH_k \tau} = P_k + \epsilon P^\perp_k =
 \exp(-i\theta_k\bmsigman)
 \pmatrix{1& 0\cr 0& \epsilon \cr}
 \exp(i\theta_k\bmsigman) \equiv P'_k ,
 \label{eq:effhamev}
 \eeq
where (inserting $\hbar$)
 \andy{epsest}
 \beq
 \epsilon \equiv e^{-V \tau/\hbar}
 \label{eq:epsest}
 \eeq
 is a parameter yielding an estimate of the efficiency of the
polarizer. One can estimate a minimal value for $V$: for a thermal
neutron (speed $v \simeq 2000$m/s) and an absorption length $\ell$ 
on the order of 1$\;$cm for the
wrong-spin component, one gets $\tau = \ell/v
\simeq 5 \mu$s and one obtains a good polarizer for $V > \hbar / \tau
\simeq 10^{-29}\;$J $\simeq 10^{-7}\;$meV.

 The evolution can be computed by using the technique of
Section~\ref{sec-potboil} ($\sqrt{P'_0}=P_0+\epsilon^{1/2}P^\perp_0$):
 \andy{projevpri}
 \beq
 \psi'(T)=  \exp(-ia\bmsigman) \sqrt{P'_0}
 \left[ \prod_{k=1}^N  B'_k \right] \up ,
 \label{eq:projevpri}
 \eeq
 with $T=N\tau$ and
 \andy{prodBpri}
 \barr
 \prod_{k=1}^N  B'_k &=&
  \prod_{k=1}^N \sqrt{P'_0} (1+i\delta \theta \bmsigman ) \sqrt{P'_0}
 =
   \prod_{k=1}^N
P'_0+i \sqrt{P'_0} (\delta \theta \bmsigman ) \sqrt{P'_0}      \nonumber\\
 & = &
 \pmatrix{1+i \delta \theta n_z& i \delta \theta \epsilon^{1/2} n_-\cr
i \delta \theta \epsilon^{1/2} n_+& \epsilon(1-i \delta \theta n_z) \cr}^N ,
  \label{eq:prodBpri}
 \earr
 where $n_\pm \equiv n_x \pm i n_y$. The evaluation of the above matrix
product when $N \to \infty$ is lengthy but straightforward. One 
gets 
 \andy{finpri}
 \beq
 \psi'(T)=\exp(-ia\bmsigman){\cal M}\phi_0,
 \label{eq:finpri}
 \eeq
 where
 \andy{fin1}
 \beq
{\cal M} =\frac{e^{-ab}}{\Delta}
 \pmatrix{\Delta\;\ch(a\Delta)+(b+in_z)\;\sh(a\Delta)& in_{-}\;\sh(a\Delta)\cr
in_{+}\;\sh(a\Delta)& \Delta\;\ch(a\Delta)-(b+in_z)\;\sh(a\Delta)\cr},
 \label{eq:fin1}
 \eeq
 with
 \andy{Deltadef}
 \beq
b = \frac{VT}{2a\hbar}, \qquad
 \Delta=\sqrt{b^2+2ibn_z-1}.
 \label{eq:Deltadef}
 \eeq
We are interested in the limit of large $b=VT/2a\hbar$. Indeed, 
larger values of $b$ correspond to more ideal polarizers. In fact 
$\gamma=V/\hbar$ represents the absorption rate of the wrong 
component of the spin, while $\omega=2a/T$ is the angular velocity 
of precession (the spin describes an angle of $2a$ in time $T$). 
The parameter $b=\gamma/\omega$ is the ratio of these two 
quantities. Large values of $b$ imply 
 \andy{gammaomega}
 \beq
 \gamma\gg \omega,
 \label{eq:gammaomega}
 \eeq
 i.e., an absorption rate much larger than the velocity of precession.
In other words, the spin rotation must be sufficiently slow to allow the
absorption of the wrong component of the spin.
By introducing the neutron speed $v$, one can
define the absorption length $\ell=v/\gamma=v\hbar/V$ and the length
covered by the neutron while rotating for $1$ rad, $L=v/\omega=vT/2a$. Hence
(\ref{eq:gammaomega}) reads
 \beq
L\gg\ell.
 \eeq 
These are all conditions of adiabaticity.

 In the large $b$ limit, using the definition (\ref{eq:Deltadef}),
(\ref{eq:fin1}) becomes
 \barr
{\cal M}
&=&\frac{e^{a(\Delta-b)}}{2\Delta}
 \pmatrix{\Delta+b+in_z& in_{-}\cr
in_{+}& \Delta-b-in_z\cr}+{\rm O}(e^{-2ab})\nonumber\\
&=&\exp(ian_z)
 \pmatrix{1-a\frac{1-n_z^2}{2b}& i\frac{n_{-}}{2b}\cr
i\frac{n_{+}}{2b}& 0\cr} +{\rm O}\left(\frac{1}{b^2}\right).
 \earr
 Remembering the definition of $b$ in (\ref{eq:Deltadef}), one gets
 \barr
{\cal M}&=&\exp(ian_z)
 \pmatrix{1+\frac{\hbar a^2(n_z^2-1)}{VT}& i\frac{\hbar an_{-}}{VT}\cr
i\frac{\hbar an_{+}}{VT}& 0\cr} +{\rm 
O}\left(\left(\frac{2a\hbar}{VT}\right)^2\right) 
 \nonumber\\
& &\longrightarrow \exp(ian_z) P_0,
 \quad\mbox{when}\quad \frac{VT}{2a\hbar}\to\infty.
 \earr
The above formula yields the first corrections to an ideal, purely 
adiabatic evolution. Basically, the system is projected on slightly 
different directions, thereby rotating in spin space. But if the 
system ``on its own" (i.e., through its dynamics) manages to rotate 
significantly between projections, then more will be absorbed on 
the next projection and it will not follow the rotating field, at 
least not without loss of probability (or intensity). 

It is interesting to note that the same result can be obtained by
considering a continuous version of the effective Hamiltonian
(\ref{eq:effham})
 \andy{effhamcont}
 \beq
H(t)=-iV P^\perp (t)=-iV U^\dagger(t) P^\perp_0 U(t),
 \label{effhamcont}
 \eeq
 where
 \andy{unrot}
 \beq
U(t)=\exp\left(i\frac{a}{T}t\;\bmsigman\right)
 \label{eq:unrot}
 \eeq
 is a unitary operator (rotation). The state vector $\psi(t)$ satisfies
the Schr\"odinger equation
 \beq
i\partial_t \psi(t)=H(t) \psi(t).
 \eeq
 Consider now the following rotated vector
 \beq
 \tltl \psi(t)=U(t) \psi(t).
 \eeq
 It is easy to prove that it satisfies the equation
 \beq
i\partial_t \tltl\psi(t)=\tltl H \tltl\psi(t),
 \eeq
where
 \beq
 \tltl H=i \dot U(t) U^\dagger(t) + U(t) H(t) U^\dagger(t)
= -\frac{a}{T}\bmsigman-i V P^\perp_0
 \eeq
is independent of $t$. One then gets
 \andy{evol}
 \beq
 \psi(t)=U^\dagger(t)\tltl\psi(t)
=\exp\left(-i\frac{a}{T}t\;\bmsigman\right)\exp(-i\tltl H t) \psi(0),
 \label{eq:evol}
 \eeq
where
 \beq
 \tltl H T=-a\bmsigman - i V T P^\perp_0=-a M,
 \quad M=\pmatrix{n_z& n_{-}\cr
n_{+}& -n_z+i2b\cr}, 
 \eeq
$b$ being defined in (\ref{eq:Deltadef}). Hence one obtains 
 \beq
 \exp(-i\tltl H T)=\exp(ia M)={\cal M}
 \eeq
and (\ref{eq:evol}) yields (\ref{eq:finpri}).  Observe that
 \beq
 \tltl H=-\omega\frac{\bmsigman}{2}-i\gamma P^\perp_0,
 \eeq
 from which it is apparent the previous interpretation of the
coefficients $\omega$ and $\gamma$.

The above calculation was performed by assuming that no external 
field is present. However, we do need an external $B$ field, in 
order to align $^3$He. Its effect can be readily taken into account 
by noticing that, when the neutron crosses the region containing 
polarized $^3$He, if the conditions for adiabaticity are satisfied, 
the neutron spin will always be (almost) parallel to the direction 
of $^3$He and therefore to the direction of the magnetic field. The 
resulting dynamical phase is therefore trivial to compute and reads 
$\phi_{\rm dyn} \simeq \mu B T/\hbar$. In order to obtain the 
geometric phase in a realistic experiment, such a dynamical phase 
should be subtracted from the total phase acquired by the neutron 
during its interaction with $^3$He. Incidentally, notice that this 
is experimentally feasible: one can take into account the 
contribution  of a large dynamical phase due to the magnetic field 
and neatly extract a small Berry phase \cite{Ioffe}. The novelty of 
our proposal consists in the introduction of polarizing $^3$He to 
force the neutron spin to follow a given trajectory is spin space. 

An alternative realization relies on a set of discrete  $^3$He
polarization filters with progressively tilted polarization axes, 
as a finite-difference approximation to the system discussed above.  
Such a system would be a neutron analog of a set of polaroid 
filters with progressively tilted axes through which a photon beam 
propagates with little or no loss (in the limit of small angles) as 
proposed by Peres \cite{Beskow}. However, in the case 
discussed in this Letter, the axes of the neutron polarizers need 
not belong to a single plane and the neutron can acquire a Berry 
phase as well as change in polarization direction. 

\vspace*{1cm}
  
 {\bf Acknowledgments:}
This work was supported in part by the United States National Science
Foundation grant PHY 97 21459.

 \end{document}